\documentclass[conference]{IEEEtran}
\usepackage{amsfonts}
\usepackage{mathrsfs}
\usepackage{amssymb,amsmath}
\usepackage{stfloats}
\usepackage[noblocks]{authblk}
\usepackage{flushend}
\usepackage{algorithm}
\usepackage{algorithmic}
\usepackage{amsmath,amssymb,epsfig,graphics,subfigure}
\usepackage{theorem}
\usepackage{array,color}
\usepackage[compress]{cite}

\makeatletter

\newcommand{\Rmnum}[1]{\expandafter\@slowromancap\romannumeral #1@}
\makeatother

\theoremheaderfont{\normalfont\bfseries}

\begin{document}

\title{Performance Analysis for Heterogeneous Cellular Systems with Range Expansion}

\author[1]{Haichuan Ding}
\author[2]{Shaodan Ma}
\author[1]{Chengwen Xing}
\author[1]{Zesong Fei}
\author[1]{Jingming Kuang}
\affil[1]{School of Information and Electronics, Beijing Institute of Technology, China \authorcr Email:\{dhcbit, xingchengwen\}@gmail.com, \{feizesong, jmkuang\}@bit.edu.cn}
\affil[2]{Department of Electrical and Computer Engineering, University of Macau, Macau, \authorcr Email:shaodanma@umac.mo}

\maketitle

\begin{abstract}
Recently heterogeneous base station structure has been adopted in cellular systems to enhance system throughput and coverage. In this paper, the uplink coverage probability for the heterogeneous cellular systems is analyzed and derived in closed-form. The randomness on the locations and number of mobile users is taken into account in the analysis. Based on the analytical results, the impacts of various system parameters on the uplink performance are investigated in detail. The correctness of the analytical results is also verified by simulation results. These analytical results can thus serve as a guidance for system design without the need of time consuming simulations.

\end{abstract}

\section{Introduction}
The ever increasing demands for high data rate and low outage probability have driven researchers around the world to keep developing new technologies for wireless communications. Heterogeneous cellular system with various tiers of base stations (BSs) is one of those new technologies proposed recently to increase coverage and system throughput \cite{Li2011}. A practical heterogenous cellular system as shown in Fig. 1 involves two tiers of BSs: macro BSs and pico BSs. With the deployment of pico BSs, the system throughput in hot-spots and/or the coverage in black holes can be significantly improved. Usually pico BSs transmit at a lower power comparing to macro BSs so that significant interference to the existing macro BSs is avoided. However, this low power transmission of pico BSs limits mobile users from connecting to them. In order to compensate this low power transmission, a power offset is generally adopted to favor the selection of pico base stations \cite{ericsson}. This scheme is called cell range expansion. Although the benefit of heterogeneous cellular systems with range expansion has been justified by some simulations, a theoretical analysis for such systems which can provide a guidance to system design is still lacking.


\begin{figure}[h]
  \begin{center}
  \includegraphics[width=2in]{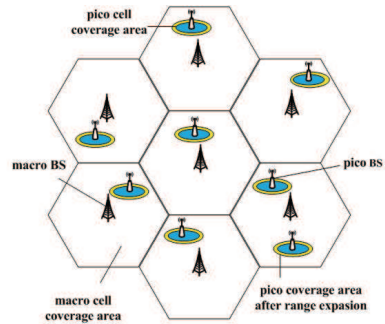}
  \end{center}
  \begin{center}
   \parbox{6cm}{\caption{A heterogenous cellular system with two tiers of base staions and range expansion.}}
  \end{center}
  \label{fig:rangeexpansion}
\end{figure}
In this paper, the uplink coverage probability for a heterogeneous cellular system with two tiers of BSs and range expansion is analyzed. Due to high mobility of mobile users and random deployment of pico BSs, the locations and numbers of pico BSs and mobile users are essentially random. Such randomness is taken into account in the performance analysis. This is significantly different from most of the analysis for cellular systems which only consider fixed topology of BSs and mobile users \cite{Wyner1994}. Based on the theory of stochastic geometry which is a powerful tool in modeling the randomness of nodes and has been adopted recently to analyze a number of wireless systems in \cite{Dhillon,Jo2011,Baccelli2011,Novlan2011,Wang201112,Wang20114,Ganti2011}, the uplink coverage probability is derived in closed-form. The correctness of the analytical results is verified by simulation results and the impacts of various system parameters on the coverage probability are discussed in detail. Specifically, the benefits of deployment of pico BSs and range expansion are theoretically justified by the analytical results. This analytical result provides an efficient way to evaluate the performance under various settings and thus provides a guidance for system design without the need of time consuming simulations.

The rest of this paper is organized as follows. System model is introduced in Section II. The probability distributions of the distances between a mobile user and its closest macro and pico BSs are analyzed in Section III. With the probability distributions of the distances, the uplink coverage probability is then derived in closed-form in Section IV. In Section V, the analytical results are verified by simulation results and the impacts of various system parameters including power offset and intensity of pico BSs are discussed in detail. Finally, conclusions are drawn in Section VI.

\section{System Model}
\begin{figure}[!ht]
  \begin{center}
  \includegraphics[width=3in]{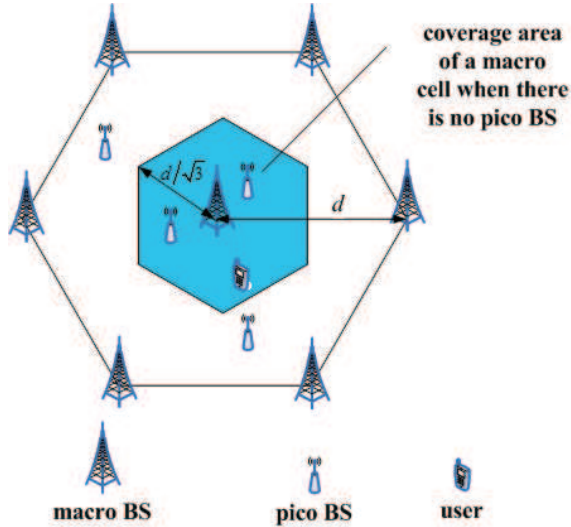}
  \end{center}
  \begin{center}
   \parbox{6cm}{\caption{A hexagonal cell with two tiers of BSs}}
  \end{center}
  \label{fig:macrobasestation}
\end{figure}
A heterogeneous cellular system with two tiers of base stations, i.e., macro and pico base stations, is considered in this paper. Generally macro BSs are placed in a planned manner and distributed deterministically. A widely used hexagonal model is thus adopted here to model the cell with macro BSs distributed at the center and vertexes as shown in Fig. 2. The radius of the hexagonal cell is denoted as $d$ and the transmission power of macro BSs is $P_1$. On the other hand, the pico BSs are usually employed to increase the system throughput in hot-spots or to improve the coverage in black holes. The number of pico BSs is variable and their locations are also irregular. It is thus reasonable to model them as a stationary Poisson point process ${\Phi ^2}$ with intensity of ${\lambda _2}$. The pico BSs generally transmit at a low transmission power ${P_2}$ with $P_2<P_1$.  Similarly, due to high mobility of mobile users, the users are assumed to form another stationary Poisson point process ${\Phi ^0}$ with intensity of ${\lambda _0}$. Without loss of generality, the mobile users are assumed to transmit with the same power as ${P_0}$.


For an arbitrary transmitter located at $z$, the channel between the transmitter and its destination located at $x$ considers both the small scale Rayleigh fading effect and path loss effect. The channel power gain is then modeled as ${h_{xz}}l(|x - z|)$ where ${h_{xz}}$ captures the Rayleigh fading effect and is modeled as an exponential random variable with unit mean, while $l(|x - z|)$ denotes the path loss and is given by  $l(|x - z|) = {\left\| {x - z} \right\|^{ - \alpha }}$ with path loss exponent of $\alpha $.  Since the path loss exponent is larger than 2 in most cases \cite{Goldsmith2005}, $\alpha  > 2$ is adopted here. The signal power received at the destination can then be formulated as ${{{h_{xz}}P_i} \mathord{\left/
 {\vphantom {{{h_{xz}}P_i} {{{\left\| {x - z} \right\|}^\alpha }}}} \right.
 \kern-\nulldelimiterspace} {{{\left\| {x - z} \right\|}^\alpha }}}$, $i=0,1,2$. Consequently, the uplink signal-to-interference-and-noise ratio (SINR) for a mobile located at $z$ and its associated BS located at $x$ is given by
 \begin{equation}
{\mbox{SINR}_z} = \frac{{{h_{xz}}{P_0}{{\left\| {x - z} \right\|}^{ - \alpha }}}}{{\sum\limits_{{z_i} \in {\Phi ^0}/\{ z\} } {{h_{x{z_i}}}{P_0}{{\left\| {x - {z_i}} \right\|}^{ - \alpha }}}  + {\sigma ^2}}},
\end{equation}
where $\sum\limits_{{z_i} \in {\Phi ^0}/\{ z\} } {{h_{x{z_i}}}{P_0}{{\left\| {x - {z_i}} \right\|}^{ - \alpha }}}$ represents the interference from other mobile users and ${{\sigma ^2}}$ is the noise power. It is generally assumed that the signal can be successfully detected/transmitted when the SINR is larger than a predermined threshold, which is a system parameter depending on modulation and coding schemes.

In the considered cellular system, each mobile can be served by either a macro or a pico BS. In practice, BS selection is performed based on the strength of the received downlink reference signals from the base stations. In order to avoid frequent handover due to temporary fading effect, the downlink reference signal strength averaged over a certain time period is taken as a metric for BS selection. It means only path loss effect will be taken into account in BS selection and the user will connect to its closest base station if BSs transmit at the same power. Denoting distances between a typical user and its closest macro and pico BSs as ${r_1}$ and ${r_2}$ respectively, the typical user will connect to the closest macro BS if ${P_1}{r_1}^{ - \alpha } \ge {P_2}{r_2}^{ - \alpha }$, otherwise it will connect to the closest pico BS. In order to alleviate the load of macro BSs and expand the range of pico BSs, a practical range expansion method \cite{ericsson} is adopted here and the BS selection is conducted based on an offset power criterion. Specifically, a typical user will connect to its closest macro BS only if ${P_1}{r_1}^{ - \alpha } \ge {P_2}\Delta {r_2}^{ - \alpha }$, where $\Delta$ is the power offset, otherwise it will connect to its closest pico BS. Since the distances of ${r_1}$ and ${r_2}$ are key factors for BS selection and uplink performance analysis, their distributions will be derived first in the next section.


\section{the distances to the closest base stations}
As stated above, the macro BSs are located at the center and vertexes of a regular hexagon as shown in Fig. 2. Due to the symmetry of the hexagon, only an equilateral triangle with macro BSs located at its vertexes (as shown in Fig. 3) is required to consider for the analysis of the distance $r_1$ between a mobile user and its closest macro BS. Since mobile users form a stationary Poisson point process, they are uniformly distributed in the considered area of the triangle. Therefore, the distance $r_1$ between a mobile user and its closest macro BS can not be larger than ${d \mathord{\left/{\vphantom {d {\sqrt 3 }}} \right.\kern-\nulldelimiterspace} {\sqrt 3 }}$ and the cumulative distribution function (CDF) of $r_1$ equals
\begin{equation}
{F_1}\left( r \right) = {\rm P}\left( {{r_1} \le r} \right) = \frac{A}{S},
\end{equation}
where $S$ is the area of the triangle OED in Fig. 3 and $A$ is the area of the intersection of the triangle OED and the union of three circles centered at the vertices of OED with radius $r$.

%

\begin{figure}[h]
  \begin{center}
  \includegraphics[width=2in]{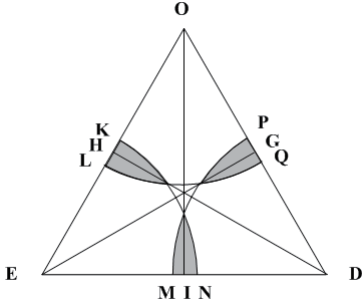}
  \end{center}
  \begin{center}
   \parbox{6cm}{\caption{Distribution analysis of $r_1$.}}
  \end{center}
  \label{fig:macrobasestation}
\end{figure}

Clearly, the area of the triangle OED (i.e., $S$) is ${{\sqrt 3 d^2} \mathord{\left/ {\vphantom {{\sqrt 3 d} 4}} \right. \kern-\nulldelimiterspace} 4}$ and the sectors OLQ, EKN and DPM will not intersect with each other when their radiuses are less than ${d \mathord{\left/
 {\vphantom {d 2}} \right.
 \kern-\nulldelimiterspace} 2}$. It is easy to obtain the area of the union of the three sectors (i.e., $A$) as $A={3 \times \frac{{\pi {r^2}}}{6}}$ when $r \leq {d \mathord{\left/{\vphantom {d 2}} \right.\kern-\nulldelimiterspace} 2}$. It follows that in this case, the probability that $r_1 \leq r$ is given by
\begin{equation}
{\rm P}\left( {{r_1} \le r} \right) = \frac{{3 \times \frac{{\pi {r^2}}}{6}}}{{\frac{{\sqrt 3 }}{4}{d^2}}} = \frac{{2\pi }}{{\sqrt 3 }}{\left( {\frac{r}{d}} \right)^2}, r \leq {d \mathord{\left/{\vphantom {d 2}} \right.\kern-\nulldelimiterspace} 2}.
\end{equation}
When ${d \mathord{\left/ {\vphantom {d 2}} \right. \kern-\nulldelimiterspace} 2} \le {r} \le {d \mathord{\left/ {\vphantom {d {\sqrt 3 }}} \right.
 \kern-\nulldelimiterspace} {\sqrt 3 }}$, the derivation of the intersecting area of $A$ is a little bit more complicated. Considering the overlapping area $B$ of the three sectors OLQ, EKN and DPM as the gray area shown in Fig. 3, after some tedious geometric calculations, we have the overlapping area as
 \begin{equation}
B = 3 \times \left(r^2\arccos \frac{d}{{2r}} - \frac{d}{2}\sqrt {r^2 - \frac{{{d^2}}}{4}}\right).
\end{equation}
It follows that when ${d \mathord{\left/ {\vphantom {d 2}} \right. \kern-\nulldelimiterspace} 2} \le {r} \le {d \mathord{\left/ {\vphantom {d {\sqrt 3 }}} \right.
 \kern-\nulldelimiterspace} {\sqrt 3 }}$, the area of the union of the three sectors (i.e., $A$) is given by
 \begin{equation}
A = {{\pi r^2} \mathord{\left/
 {\vphantom {{\pi r^2} 2}} \right.
 \kern-\nulldelimiterspace} 2} - 3 \times \left(r^2\arccos \frac{d}{{2r}} - \frac{d}{2}\sqrt {r^2 - \frac{{{d^2}}}{4}}\right).
\end{equation}
%
%
Therefore, the probability of $r_1 \leq r$ when ${d \mathord{\left/
 {\vphantom {d 2}} \right.
 \kern-\nulldelimiterspace} 2} \le {r} \le {d \mathord{\left/
 {\vphantom {d {\sqrt 3 }}} \right.
 \kern-\nulldelimiterspace} {\sqrt 3 }}$ can be expressed as
\begin{align}
{\rm P}\left( {{r_1} \le r} \right) &= \frac{{2\pi }}{{\sqrt 3 }}{\left( {\frac{r}{d}} \right)^2} - 4\sqrt 3 {\left( {\frac{r}{d}} \right)^2}\arccos \frac{d}{{2r}} \nonumber \\ &+ \frac{{\sqrt {12{r^2} - 3{d^2}} }}{d}.
\end{align}
Finally, the CDF of $r_1$ can be written as (\ref{F_r_1}) as shown on the top of next page.
\begin{figure*}[ht]
\begin{equation}
\label{F_r_1}
{F_1}({r}) = \left\{ {\begin{array}{*{20}{c}}
{\frac{{2\pi }}{{\sqrt 3 }}{{\left( {\frac{r}{d}} \right)}^2}}&{0 < {r} \le \frac{d}{2}}\\
{\frac{{2\pi }}{{\sqrt 3 }}{{\left( {\frac{r}{d}} \right)}^2} - \left( {4\sqrt 3 {{\left( {\frac{r}{d}} \right)}^2}\arccos \frac{d}{{2r}} - \frac{{\sqrt {12{r^2} - 3{d^2}} }}{d}} \right)}&{\frac{d}{2} < {r} \le \frac{d}{{\sqrt 3 }}}
\end{array}} \right.
\end{equation}
\end{figure*}

With respect to the distance of $r_2$ between a mobile user and its closest pico BS, the probability that ${{r_2} \le r}$ equals the probability that there are no pico BSs in the circle centered at the user with radius ${r}$. The CDF of ${r_2}$ can be derived as follows
\begin{equation}
\label{F_r_2}
{F_2}({r}) = {\rm P}\left( {{r_2} \le r} \right) = 1 - {\rm{P}}\{ {\Phi^2} (B(z,{r})) = 0\}  = 1 - {e^{ - {\lambda _2}\pi {r}^2}},
\end{equation}
where $z$ is the location of the mobile user, $B(z,{r})$ is a circle centered at $z$ with a radius of $r$ and ${\Phi^2} (B(z,{r}))$ denotes the number of pico BSs in $B(z,{r})$.

In the following section, these distributions will be used to analyze the uplink performance of the considered heterogenous cellular systems with macro and pico BSs.

\section{uplink performance analysis}

In this section, the coverage probability of the considered uplink heterogeneous cellular system, which is defined as the probability that a mobile user can successfully communicate with its associated BS, is analyzed. In fact, this coverage probability is the complementary of the outage probability of a mobile user and has been widely adopted as a performance metric for system evaluation \cite{Dhillon,Baccelli2011,Baszczyszyn2009}. Since each mobile user could access to either its closest macro BS or to its closest pico BS, the uplink coverage probability ${{\rm P}_c}$ can be written as the sum of two probabilities: the probability ${{\rm P}_{msuc}}$ that the user connects to its closest macro BS and conducts a successful transmission and the probability ${{\rm P}_{psuc}}$ that a user connects to its closest pico BS and conducts a successful transmission, namely


\begin{equation}
\label{P_c}
{{\rm P}_c} = {{\rm P}_{msuc}} + {{\rm P}_{psuc}}.
\end{equation}
Here a successful transmission means the signal can be successfully decoded by the receiver, i.e., the corresponding SINR is larger than a predefined threshold. In the following, the two probabilities ${{\rm P}_{msuc}}$ and ${{\rm P}_{psuc}}$ will be derived individually.

According to the definition, ${{\rm P}_{msuc}}$ can be formulated as
\begin{align}
\label{macro_coverage}
{{\rm{P}}_{msuc}} = {{\rm{E}}_{{r_1},{r_2}}}\left[ {1\left( {{r_2} \ge {{\left( {\frac{{{P_2}\Delta }}{{{P_1}}}} \right)}^{\frac{1}{\alpha }}}{r_1}} \right)1\left( {{\gamma _1} \ge {T_1}} \right)} \right],
\end{align}
where $T_1$ is the threshold above which macro BSs could decode successfully, and ${{\gamma _1}}$ is the received SINR at the user's closest macro BS and equals
\begin{equation}
\label{SINR_1}
{\gamma _1} = \frac{{h_1{P_0}l({r_1})}}{{I_1 + N}},
\end{equation}
where $N$ is the additive noise power, $h_1$ denotes the exponentially distributed channel power gain between the user and its closest macro BS, and $I_1$ represents the received interference as
\begin{equation}
\label{Interference_1}
{I_1} = \sum\limits_{{z_i} \in {\Phi ^0}/\{ z\} } {{h_{{xz_i}}}{P_0}l\left( {|x-{z_i}|} \right)}.
\end{equation}
In (\ref{Interference_1}), $z$ represents the location of the typical user, ${{\Phi ^0}/\{ z\} }$ is the set formed by the points of ${\Phi ^0}$ except $z$, and $x$ denotes the location of the closest macro BS concerned. Since the event ${{r_2} \ge {{\left( {\frac{{{P_2}\Delta }}{{{P_1}}}} \right)}^{\frac{1}{\alpha }}}{r_1}}$ is independent with the event ${\gamma _1} \ge {T_1}{\rm{ }}$, the probability corresponding to the closest macro BS in (\ref{macro_coverage}) can be rewritten as
 \begin{equation}
 \label{macro_coverage_P}
{{\rm{P}}_{msuc}} = {{\rm{E}}_{{r_1}}}\left[ {{\rm{P}}\left( {{r_2} \ge {{\left( {\frac{{{P_2}\Delta }}{{{P_1}}}} \right)}^{\frac{1}{\alpha }}}{r_1}} \vert {r_1} \right){\rm{P}}\left( {{\gamma _1} \ge {T_1}} \vert {r_1} \right)} \right].
 \end{equation}
Based on (\ref{SINR_1}), ${\rm{P}}\left( {{\gamma _1} \ge {T_1}} \vert {r_1} \right)$ can be reformulated as
\begin{align}
{\rm{P}}\left( {{\gamma _1} \ge {T_1}} \vert {r_1} \right) &= {\rm P}\left( {h_1 \ge \frac{{{T_1}\left( {{I_1} + N} \right)}}{{{P_0}l\left( {{r_1}} \right)}}} \vert {r_1} \right) \nonumber \\ & = {\rm E_{I_1}}\left[ {{e^{ - \frac{{{T_1}\left( {{I_1} + N} \right)}}{{{P_0}l\left( {{r_1}} \right)}}}}} \right],
\end{align}
Since cellular networks are generally interference dominated, the additive noise $N$ can be neglected. Based on (\ref{Interference_1}) and Slivnyak Theorem regarding to the interference \cite{Baszczyszyn2009}, we can get
\begin{equation}
\label{P_SINR1}
{\rm{P}}\left( {{\gamma _1} \ge {T_1}|{r_1}} \right) = {e^{ - \frac{{2{\pi ^2}{\lambda _0}}}{{\alpha \sin \left( {{{2\pi } \mathord{\left/
 {\vphantom {{2\pi } \alpha }} \right.
 \kern-\nulldelimiterspace} \alpha }} \right)}}{T_1}^{\frac{2}{\alpha }}r_1^2}}.
\end{equation}
Meanwhile, we have
\begin{align}
\label{P_r2}
{\rm{P}}\left( {{r_2} \ge {{\left( {\frac{{{P_2}\Delta }}{{{P_1}}}} \right)}^{\frac{1}{\alpha }}}{r_1}} \vert {r_1} \right) &= \int_{{{\left( {{{{P_2}\Delta } \mathord{\left/
 {\vphantom {{{P_2}\Delta } {{P_1}}}} \right.
 \kern-\nulldelimiterspace} {{P_1}}}} \right)}^{{1 \mathord{\left/
 {\vphantom {1 \alpha }} \right.
 \kern-\nulldelimiterspace} \alpha }}}{r_1}}^\infty  {d{F_2}\left( {{r_2}} \right)} \nonumber \\ & = {e^{ - \pi {\lambda _2}{{\left( {\frac{{{P_2}\Delta }}{{{P_1}}}} \right)}^{\frac{2}{\alpha }}}r_1^2}},
\end{align}
Putting (\ref{P_SINR1}) and (\ref{P_r2}) into (\ref{macro_coverage_P}), we have
\begin{align}
\label{P_msuc_1}
{{\rm{P}}_{msuc}} = \int\limits_0^{\frac{d}{{\sqrt 3 }}} {{e^{ - \left( {\frac{{2{\pi ^2}{\lambda _0}}}{{\alpha \sin \left( {{{2\pi } \mathord{\left/
 {\vphantom {{2\pi } \alpha }} \right.
 \kern-\nulldelimiterspace} \alpha }} \right)}}{T_1}^{\frac{2}{\alpha }} + \pi {\lambda _2}{{\left( {\frac{{{P_2}\Delta }}{{{P_1}}}} \right)}^{\frac{2}{\alpha }}}} \right)r_1^2}}d{F_1}({r_1})},
\end{align}
Using integration by parts and with the CDF of $r_1$ in (\ref{F_r_1}), (\ref{P_msuc_1}) can be derived as


\begin{align}
\label{P_msuc_3}
{{\rm{P}}_{msuc}} &={e^{ - K\frac{{{d^2}}}{3}}} - \int\limits_0^{\frac{d}{{\sqrt 3 }}} {\frac{{4K\pi }}{{\sqrt 3 {d^2}}}r_1^3{e^{ - Kr_1^2}}d{r_1}}  \nonumber \\ & - \int\limits_0^{\frac{d}{{\sqrt 3 }}} {\frac{{2K}}{d}\sqrt {12r_1^2 - 3{d^2}} {r_1}{e^{ - Kr_1^2}}d{r_1}} \nonumber \\ & + \int\limits_0^{\frac{d}{{\sqrt 3 }}} {\frac{{8\sqrt 3 K}}{{{d^2}}}{r_1}^3\arccos \frac{d}{{2{r_1}}}{e^{ - Kr_1^2}}d{r_1}},
\end{align}
where $K$ equals
\begin{equation}
K{\rm{ = }}\frac{{2{\pi ^2}{\lambda _0}}}{{\alpha \sin \left( {{{2\pi } \mathord{\left/
 {\vphantom {{2\pi } \alpha }} \right.
 \kern-\nulldelimiterspace} \alpha }} \right)}}{T_1}^{\frac{2}{\alpha }} + \pi {\lambda _2}{\left( {\frac{{{P_2}\Delta }}{{{P_1}}}} \right)^{\frac{2}{\alpha }}}.
\end{equation}
Based on 2.322 in \cite{Gradshteyn2007}, the second term in the right hand side of (\ref{P_msuc_3}) can be rewritten as
\begin{align}
\label{Second_term}
\int\limits_0^{\frac{d}{{\sqrt 3 }}} {\frac{{4K\pi }}{{\sqrt 3 {d^2}}}r_1^2{e^{ - Kr_1^2}}d{r_1}{\rm{ = }}\frac{{{\rm{2}}\pi }}{{\sqrt 3 {d^2}K}}\left( {1 - {e^{ - K\frac{{{d^2}}}{3}}}} \right)}  - \frac{{{\rm{2}}\pi }}{{3\sqrt 3 }}{e^{ - K\frac{{{d^2}}}{3}}}.
\end{align}
Similarly, with 3.382 in \cite{Gradshteyn2007}, the third term in the right hand side of (\ref{P_msuc_3}) can be derived as
\begin{align}
\
&\int\limits_0^{\frac{d}{{\sqrt 3 }}} {\frac{{2K}}{d}\sqrt {12r_1^2 - 3{d^2}} {r_1}{e^{ - Kr_1^2}}d{r_1}}  =\nonumber \\ & \sqrt {\frac{{12}}{{K{d^2}}}} {e^{ - K\frac{{{d^2}}}{4}}}\left( {\Gamma \left( {\frac{3}{2}}, 0 \right) - \Gamma \left( {\frac{3}{2},\frac{{K{d^2}}}{{12}}} \right)} \right),
\end{align}
where $\Gamma \left( . , . \right)$ is the upper incomplete gamma function defined as $\Gamma \left( {c,x} \right) = \int\limits_x^\infty  {{e^{ - t}}{t^{c - 1}}dt}$. With the properties that $\Gamma \left( {c + 1,x} \right) = c\Gamma \left( {c,x} \right) + {x^c}{e^{ - x}}$ and $\Gamma \left( {\frac{1}{2},{x^2}} \right) = \sqrt \pi-\sqrt \pi  \left(1-2{Q\left( {\sqrt 2 x} \right)}\right)$, where $Q (.)$ is the $Q$ function, the third term can be further given as


\begin{align}
\label{Third_term}
&\int\limits_0^{\frac{d}{{\sqrt 3 }}} {\frac{{2K}}{d}\sqrt {12r_1^2 - 3{d^2}} {r_1}{e^{ - Kr_1^2}}d{r_1}}  =\nonumber \\& \sqrt {\frac{{3\pi }}{{K{d^2}}}} {e^{ - K\frac{{{d^2}}}{4}}}\left(1- 2Q \left( {\sqrt {\frac{{K{d^2}}}{{6}}} } \right)\right) - {e^{ - K\frac{{{d^2}}}{3}}}.
\end{align}
Substituting (\ref{Second_term}) and (\ref{Third_term}) into (\ref{P_msuc_3}) and after tedious computation, we finally can get the probability corresponding to the closest macro BS, ${{\rm P}_{msuc}}$, as (\ref{P_msuc_Final}) as shown on the top of next page, where ${{{\rm{P'}}}_G} = \int\limits_{\frac{d}{2}}^{\frac{d}{{\sqrt 3 }}} {r_1^3\arccos \frac{d}{{2{r_1}}}} {e^{ - \left( {{\lambda _2}\pi {{\left( {\frac{{{P_2}\Delta }}{{{P_1}}}} \right)}^{\frac{2}{\alpha }}} + \frac{{2{\pi ^2}{\lambda _0}{T_1}^{\frac{2}{\alpha }}}}{{\alpha \sin \left( {{{2\pi } \mathord{\left/
 {\vphantom {{2\pi } \alpha }} \right.
 \kern-\nulldelimiterspace} \alpha }} \right)}}} \right)r_1^2}}d{r_1}$.

\begin{figure*}[ht]
\begin{align}
\label{P_msuc_Final}
{{\rm P}_{msuc}}& = \frac{{2\pi \left( {1 - {e^{ - \left( {\frac{{2{\pi ^2}{\lambda _0}}}{{\alpha \sin \left( {{{2\pi } \mathord{\left/
 {\vphantom {{2\pi } \alpha }} \right.
 \kern-\nulldelimiterspace} \alpha }} \right)}}{{{T_1}}^{\frac{2}{\alpha }}} + \pi {\lambda _2}{{\left( {\frac{{{P_2}\Delta}}{{{P_1}}}} \right)}^{\frac{2}{\alpha }}}} \right)\frac{{{d^2}}}{3}}}} \right)}}{{\sqrt 3 {d^2}\left( {\frac{{2{\pi ^2}{\lambda _0}}}{{\alpha \sin \left( {{{2\pi } \mathord{\left/
 {\vphantom {{2\pi } \alpha }} \right.
 \kern-\nulldelimiterspace} \alpha }} \right)}}{{{T_1}}^{\frac{2}{\alpha }}} + \pi {\lambda _2}{{\left( {\frac{{{P_2}\Delta}}{{{P_1}}}} \right)}^{\frac{2}{\alpha }}}} \right)}} - \frac{{2\pi }}{{3\sqrt 3 }}{e^{ - \left( {\frac{{2{\pi ^2}{\lambda _0}}}{{\alpha \sin \left( {{{2\pi } \mathord{\left/
 {\vphantom {{2\pi } \alpha }} \right.
 \kern-\nulldelimiterspace} \alpha }} \right)}}{{{T_1}}^{\frac{2}{\alpha }}} + \pi {\lambda _2}{{\left( {\frac{{{P_2}\Delta}}{{{P_1}}}} \right)}^{\frac{2}{\alpha }}}} \right)\frac{{{d^2}}}{3}}}\nonumber \\ &+ \sqrt {\frac{{3\pi }}{{{d^2}\left( {\frac{{2{\pi ^2}{\lambda _0}}}{{\alpha \sin \left( {{{2\pi } \mathord{\left/
 {\vphantom {{2\pi } \alpha }} \right.
 \kern-\nulldelimiterspace} \alpha }} \right)}}{{{T_1}}^{\frac{2}{\alpha }}} + \pi {\lambda _2}{{\left( {\frac{{{P_2}\Delta}}{{{P_1}}}} \right)}^{\frac{2}{\alpha }}}} \right)}}} \times {e^{ - \left( {\frac{{2{\pi ^2}{\lambda _0}}}{{\alpha \sin \left( {{{2\pi } \mathord{\left/
 {\vphantom {{2\pi } \alpha }} \right.
 \kern-\nulldelimiterspace} \alpha }} \right)}}{{{T_1}}^{\frac{2}{\alpha }}} + \pi {\lambda _2}{{\left( {\frac{{{P_2}\Delta}}{{{P_1}}}} \right)}^{\frac{2}{\alpha }}}} \right)\frac{{{d^2}}}{4}}}\nonumber \\ & \times \left( {1 - 2Q\left( {\sqrt {\frac{{\left( {\frac{{2{\pi ^2}{\lambda _0}}}{{\alpha \sin \left( {{{2\pi } \mathord{\left/
 {\vphantom {{2\pi } \alpha }} \right.
 \kern-\nulldelimiterspace} \alpha }} \right)}}{T_1}^{\frac{2}{\alpha }} + \pi {\lambda _2}{{\left( {\frac{{{P_2}\Delta }}{{{P_1}}}} \right)}^{\frac{2}{\alpha }}}} \right){d^2}}}{{6}}} } \right)} \right)  - \frac{{8\sqrt 3 }}{{{d^2}}}\left( {\frac{{2{\pi ^2}{\lambda _0}}}{{\alpha \sin \left( {{{2\pi } \mathord{\left/
 {\vphantom {{2\pi } \alpha }} \right.
 \kern-\nulldelimiterspace} \alpha }} \right)}}{{{T_1}}^{\frac{2}{\alpha }}} + \pi {\lambda _2}{{\left( {\frac{{{P_2}\Delta}}{{{P_1}}}} \right)}^{\frac{2}{\alpha }}}} \right){\rm P}_G^{'},
\end{align}
\end{figure*}

Similarly, the probability that a user connects to its closest pico BS and conducts a successful transmission is given by
\begin{equation}
{{\rm{P}}_{psuc}} = {{\rm{E}}_{{r_1},{r_2}}}\left[ {1\left( {{r_2} < {{\left( {\frac{{{P_2}\Delta }}{{{P_1}}}} \right)}^{\frac{1}{\alpha }}}{r_1}} \right)1\left( {{\gamma _2} \ge {T_2}} \right)} \right],
\end{equation}
where $T_2$ is the threshold above which the pico BS can successfully decode the user's signal, and ${{\gamma _2}}$ is the received SINR at the user's closest pico BS and defined as
\begin{equation}
\label{SINR_2}
{\gamma _2} = \frac{{h_2{P_0}l({r_2})}}{{{I_2} + N}}.
\end{equation}
In (\ref{SINR_2}), ${h_2}$ denotes the channel power gain between the user and its closest pico BS, and $I_2$ is the interference seen at the pico BS and defined as
\begin{equation}
{I_2} = \sum\limits_{{z_i} \in {\Phi ^0}/\{ z\} } {{h_{{xz_i}}}{P_0}l\left( {|x-{z_i}|} \right)},
\end{equation}
where $z$ represents the location of the typical user and ${{\Phi ^0}/\{ z\} }$ is the set formed by the points of ${\Phi ^0}$ except $z$, while $x$ denotes the location of the closest pico BS. Under the interference dominated cellular system and based on Slivnyak Theorem \cite{Baszczyszyn2009}, we can get
\begin{equation}
\label{P_SINR2}
{\rm{P}}\left( {{\gamma _2} \ge {T_2}} \vert {r_2} \right) = {e^{ - \frac{{2{\pi ^2}{\lambda _0}}}{{\alpha \sin \left( {{{2\pi } \mathord{\left/
 {\vphantom {{2\pi } \alpha }} \right.
 \kern-\nulldelimiterspace} \alpha }} \right)}}{T_2}^{\frac{2}{\alpha }}r_2^2}}.
\end{equation}
With (\ref{P_SINR2}), the probability ${{\rm{P}}_{psuc}}$ can be written as
\begin{align}
{{\rm{P}}_{psuc}} = \int\limits_0^{\frac{d}{{\sqrt 3 }}} {\int\limits_0^{\sqrt[\alpha ]{{\frac{{{P_2}\Delta }}{{{P_1}}}}}{r_1}} {{e^{ - \frac{{2{\pi ^2}{\lambda _0}}}{{\alpha \sin \left( {{{2\pi } \mathord{\left/
 {\vphantom {{2\pi } \alpha }} \right.
 \kern-\nulldelimiterspace} \alpha }} \right)}}{T_2}^{\frac{2}{\alpha }}r_2^2}}d{F_2}\left( {{r_2}} \right)} d{F_1}\left( {{r_1}} \right)} .
\end{align}
Based on (\ref{F_r_1}) and (\ref{F_r_2}), and following a similar derivation as that for ${{\rm P}_{msuc}}$, the probability ${{\rm P}_{psuc}}$ can eventually be derived as (\ref{P_psuc_Final}), where ${{{\rm{P''}}}_G} = \int\limits_{\frac{d}{2}}^{\frac{d}{{\sqrt 3 }}} {r_1^3\arccos \frac{d}{{2{r_1}}}} {e^{ - \pi {{\left( {\frac{{{P_2}\Delta }}{{{P_1}}}} \right)}^{\frac{2}{\alpha }}}\left( {{\lambda _2} + \frac{{2\pi {\lambda _0}{T_2}^{\frac{2}{\alpha }}}}{{\alpha \sin \left( {{{2\pi } \mathord{\left/
 {\vphantom {{2\pi } \alpha }} \right.
 \kern-\nulldelimiterspace} \alpha }} \right)}}} \right)r_1^2}}d{r_1}$.
\begin{figure*}[ht]
\begin{align}
\label{P_psuc_Final}
{{\rm P}_{psuc}}& = \frac{{\alpha \sin \left( {{{2\pi } \mathord{\left/
 {\vphantom {{2\pi } \alpha }} \right.
 \kern-\nulldelimiterspace} \alpha }} \right){\lambda _2}}}{{2\pi {\lambda _0}{{\left( {\frac{{{T_2}}}{G}} \right)}^{\frac{2}{\alpha }}} + {\lambda _2}\alpha \sin \left( {{{2\pi } \mathord{\left/
 {\vphantom {{2\pi } \alpha }} \right.
 \kern-\nulldelimiterspace} \alpha }} \right)}} -  \frac{{2{\lambda _2}{{\left( {\frac{{{P_1}}}{{{P_2}\Delta }}} \right)}^{\frac{2}{\alpha }}}{\alpha ^2}{{\sin }^2}({{2\pi } \mathord{\left/
 {\vphantom {{2\pi } \alpha }} \right.
 \kern-\nulldelimiterspace} \alpha })}}{{\sqrt 3 {d^2}{{\left( {{\lambda _2}\alpha \sin ({{2\pi } \mathord{\left/
 {\vphantom {{2\pi } \alpha }} \right.
 \kern-\nulldelimiterspace} \alpha }) + 2\pi {\lambda _0}{{\left( {\frac{{{T_2}}}{G}} \right)}^{\frac{2}{\alpha }}}} \right)}^2}}} \times\left( {1 - {e^{ - \frac{{\pi {{\left( {\frac{{{P_2}\Delta }}{{{P_1}}}} \right)}^{\frac{2}{\alpha }}}{d^2}}}{3}\left( {{\lambda _2} + \frac{{2\pi {\lambda _0}{{\left( {\frac{{{T_2}}}{G}} \right)}^{\frac{2}{\alpha }}}}}{{\alpha \sin \left( {{{2\pi } \mathord{\left/
 {\vphantom {{2\pi } \alpha }} \right.
 \kern-\nulldelimiterspace} \alpha }} \right)}}} \right)}}} \right) \nonumber \\ &+\frac{{2\pi {\lambda _2}\alpha \sin ({{2\pi } \mathord{\left/
 {\vphantom {{2\pi } \alpha }} \right.
 \kern-\nulldelimiterspace} \alpha })}}{{3\sqrt 3 \left( {2\pi {\lambda _0}{{\left( {\frac{{{T_2}}}{G}} \right)}^{\frac{2}{\alpha }}} + {\lambda _2}\alpha \sin ({{2\pi } \mathord{\left/
 {\vphantom {{2\pi } \alpha }} \right.
 \kern-\nulldelimiterspace} \alpha })} \right)}} \times{e^{ - \frac{{\pi {{\left( {\frac{{{P_2}\Delta }}{{{P_1}}}} \right)}^{\frac{2}{\alpha }}}{d^2}}}{3}\left( {{\lambda _2} + \frac{{2\pi {\lambda _0}{{\left( {\frac{{{T_2}}}{G}} \right)}^{\frac{2}{\alpha }}}}}{{\alpha \sin \left( {{{2\pi } \mathord{\left/
 {\vphantom {{2\pi } \alpha }} \right.
 \kern-\nulldelimiterspace} \alpha }} \right)}}} \right)}}  - \frac{{\sqrt 3 {\lambda _2}{\alpha ^{\frac{3}{2}}}{{\sin }^{\frac{3}{2}}}({{2\pi } \mathord{\left/
 {\vphantom {{2\pi } \alpha }} \right.
 \kern-\nulldelimiterspace} \alpha }){{\left( {\frac{{{P_1}}}{{{P_2}\Delta }}} \right)}^{\frac{1}{\alpha }}}}}{{d{{\left( {2\pi {\lambda _0}{{\left( {\frac{{{T_2}}}{G}} \right)}^{\frac{2}{\alpha }}} + {\lambda _2}\alpha \sin ({{2\pi } \mathord{\left/
 {\vphantom {{2\pi } \alpha }} \right.
 \kern-\nulldelimiterspace} \alpha })} \right)}^{\frac{3}{2}}}}}\nonumber \\ &\times{e^{ - \frac{{\pi {{\left( {\frac{{{P_2}\Delta }}{{{P_1}}}} \right)}^{\frac{2}{\alpha }}}{d^2}}}{4}\left( {{\lambda _2} + \frac{{2\pi {\lambda _0}{{\left( {\frac{{{T_2}}}{G}} \right)}^{\frac{2}{\alpha }}}}}{{\alpha \sin \left( {{{2\pi } \mathord{\left/
 {\vphantom {{2\pi } \alpha }} \right.
 \kern-\nulldelimiterspace} \alpha }} \right)}}} \right)}}  \times\left( {1 - 2Q\left( {d{{\left( {\frac{{{P_2}\Delta }}{{{P_1}}}} \right)}^{\frac{1}{\alpha }}}\sqrt {\frac{{\pi \left( {2\pi {\lambda _0}{{\left( {\frac{{{T_2}}}{G}} \right)}^{\frac{2}{\alpha }}} + {\lambda _2}\alpha \sin \left( {{{2\pi } \mathord{\left/
 {\vphantom {{2\pi } \alpha }} \right.
 \kern-\nulldelimiterspace} \alpha }} \right)} \right)}}{{6\alpha \sin \left( {{{2\pi } \mathord{\left/
 {\vphantom {{2\pi } \alpha }} \right.
 \kern-\nulldelimiterspace} \alpha }} \right)}}} } \right)} \right) \nonumber \\ & -\frac{{24\pi {\lambda _2}}}{{\sqrt 3 {d^2}}}{\left( {\frac{{{P_2}\Delta }}{{{P_1}}}} \right)^{\frac{2}{\alpha }}}{\rm P}_G^{''},
\end{align}
\end{figure*}

Putting (\ref{P_msuc_Final}) and (\ref{P_psuc_Final}) into (\ref{P_c}), the coverage probability can then be directly obtained.

\section{Numerical results and discussion}
In this section, computer simulations are conducted to verify the correctness of the analytical results and investigate the performance of the two-tiers heterogenous cellular system. Simulation parameters are set as shown in Table I.
\begin{table}[!h]
\label{system_setting}
\begin{center}
\caption{simulation settings}
\begin{tabular}{|c|c|}
\hline
$\lambda_2$ & $3.06 \times {10^{ - 3}}$  \\
\hline
$\lambda_0$  & $7.66 \times {10^{ - 3}}$   \\
\hline
${T_1}$ & $1$   \\
\hline
${T_2}$ & $1$   \\
\hline
$d$  & ${{50} \mathord{\left/
 {\vphantom {{50} {\sqrt 3 }}} \right.
 \kern-\nulldelimiterspace} {\sqrt 3 }}$   \\
\hline
${P_1}/{P_2}$ & $100$   \\
\hline
\end{tabular}

\end{center}
\end{table}

Fig. 4 shows the coverage probability versus the power offset $\Delta$. It is clear that the analytical results match well with the simulation results, demonstrating the correctness of the analytical results. Furthermore, the coverage probability increases with the power offset and achieves its maximum when the power offset equals to ${P_1}/{P_2}$, i.e., $20dB$. The coverage probability drops slightly when the power offset is larger than $20dB$. It means that in order to achieve the maximum uplink coverage probability, the power offset is suggested to be set equal to the ratio between the transmission powers of macro and pico BSs. This is not surprised since under this setting, the pico BSs deployed for coverage improvement can serve the mobile users with equal probability as the macro BS.


\begin{figure}[!ht]
  \begin{center}
  \includegraphics[width=2in]{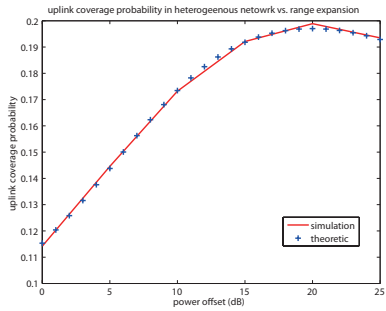}
  \end{center}
  \begin{center}
   \parbox{6cm}{\caption{uplink coverage probability vs. power offset.}}
  \end{center}
  \label{fig:macrobasestation}
\end{figure}

The effect of the intensity of pico BSs on the coverage probability is next investigated and the results are shown in Fig. 5. Here the x axis in the figure denotes the intensity of pico BSs normalized by $\lambda_2$ shown in Table I. It is shown that the uplink coverage probability increases with the intensity of pico BSs as expected. This result demonstrates the benefit of pico deployment for the cellular coverage improvement.

\section{Conclusions}
In this paper, an analytical approach to performance analysis for heterogenous cellular systems was proposed based on the theory of stochastic geometry. The uplink coverage probability was derived in closed-form and its correctness was verified by simulation results. From the analytical results, the impacts of both pico deployment and range expansion on the system's uplink performance were investigated. It has been found that in order to achieve the maximum uplink coverage probability, the power offset should be set equal to the ratio between the powers of macro and pico BSs. Meanwhile, the results have theoretically justified the benefit of the deployment of pico BSs.


\begin{thebibliography}{99}
{   \bibitem{Li2011}
    H. Li, J. Hajipour, A. Attar and V. C. M. Leung,``Efficient HETNET implementation using broadband wireless access with fiber-connected massively distributed antennas architecture,'' {\em IEEE Wireless Communications}, vol. 18, no. 3, pp. 72--78, Jun., 2011.
    \bibitem{ericsson}
    S. Landstr\"om, A. Furusk\~ar, K. Johansson, L. Falconetti and F. Kronestedt,``Heterogeneous networks increasing cellular capacity,'' {\em  Ericsson Review, Technology}, Feb. 2011.
    \bibitem{Wyner1994}
    A. D. Wyner,``Shannon-theoretic approach to a Gaussian cellular multiple-access channel,'' {\em IEEE Trans. on Info. Theory}, vol. 40, no. 6, pp. 1713--1727, Nov. 1994.
        \begin{figure}[!ht]
  \begin{center}
  \includegraphics[width=2in]{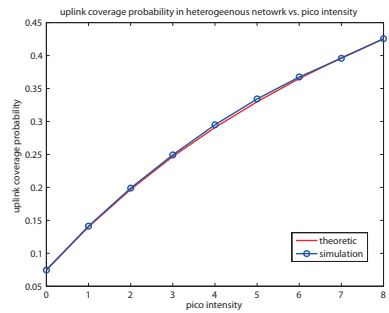}
  \end{center}
  \begin{center}
   \parbox{6cm}{\caption{uplink coverage probability vs. pico intensity.}}
  \end{center}
  \label{fig:macrobasestation}
\end{figure}
    \bibitem{Dhillon}
    H. S. Dhillon, R. K. Ganti, F. Baccelli and J. G. Andrews,`` Modeling and analysis of K-Tier downlink heterogeneous cellular networks,'' {\em IEEE J. Select. Areas Commun.}, vol. 30, no. 3, pp. 550--560, Mar. 2012.
    \bibitem{Jo2011}
    H.-S. Jo, Y. J. Sang, P. Xia, J. G. Andrews, ``Outage probability for heterogeneous cellular networks with biased cell association,'' {\em in Proc. 2011 IEEE Global Telecommunications Conference (GLOBECOM 2011)}, pp. 1--5, 2011.
    \bibitem{Baccelli2011}
    J. G. Andrews, F. Baccelli, and R. K. Ganti, ``A tractable approach to coverage and rate in cellular networks'', {\em IEEE Trans. Commun.}, vol. 59, no. 11, pp. 3122--3134, Nov. 2011.
    \bibitem{Novlan2011}
    T. D. Novlan, R. K. Ganti, A. Ghosh, and J. G. Andrews, ``Analytical evaluation of fractional frequency reuse for OFDMA cellular networks,'' {\em IEEE Trans. Wireless Commun.}, vol. 10, no. 12, pp. 4294--4350, Dec. 2011.
    \bibitem{Wang201112}
    H. Wang, S. Ma, T. Ng and H. V. Poor, "A general analytical approach for opportunistic cooperative systems with spatially random relays", {\em IEEE Trans. on Wireless Commun.}, vol. 10, no. 12, pp. 4122--4129, Dec. 2011.
    \bibitem{Wang20114}
    H. Wang, S. Ma and T. Ng, ``On performance of cooperative communications system with spatially random relays,'' {\em IEEE Trans. on Commun.}, vol. 59, no. 4, pp. 1190-1199, Apr. 2011.
    \bibitem{Ganti2011}
    R. K. Ganti, M. Haenggi, ``Spatial analysis of opportunistic downlink relaying in a Two-Hop cellular system,'' available at http://arxiv.org/abs/0911.2948.
    \bibitem{Goldsmith2005}
    A. Goldsmith, {\em Wireless Communications}, Cambridge: Cambridge press, 2005.
    \bibitem{Baszczyszyn2009}
    F. Baccelli, B. B{\l}aszczyszyn, {\em Stochastic Geometry and Wireless Networks (volume 2)}, availabe at http://hal.inria.fr/docs/00/43/87/70/PDF/FnT2.pdf.
    \bibitem{Gradshteyn2007}
    I. S. Gradshteyn and I. M. Ryzhik, {\em Tables of Integrals, Series, and Products}, San Diego: Academic press, 2007.
    \bibitem{Haenggi2009}
    M. Haenggi, J. G. Andrews, F. Baccelli, O. Dousse and M. Franceschetti, ``Stochastic geometry and random graphs for the analysis and design of wireless networks,'' {\em IEEE J. Select. Areas Commun.}, vol. 27, no. 7, pp. 1029--1046, Sep. 2009.
    \bibitem{Stoyan1995}
    D. Stoyan, W. Kendall and J. Mecke, {\em Stochastic Geometry and its Applications}, Chichester: WILEY, 1995.
}
\end{thebibliography}
\end{document}